\documentclass[aps,twocolumn,reprint,letterpaper]{revtex4-1}

\usepackage{graphicx}     

\newcommand{\h}{L}

\newcommand{\Eq}[1]{Eq.~(\ref{eq:#1})}

\begin{document}

\title{Surface van der Waals Forces in a Nutshell}

\author{Luis G. MacDowell}
\affiliation{Departamento de Qu\'{\i}mica F\'{\i}sica, Facultad de Ciencias
Qu\'{\i}micas, Universidad Complutense de Madrid, Madrid, 28040, Spain.}
\email{lgmac@quim.ucm.es}


\begin{abstract}
Most often in chemical physics,  long range van der Waals surface interactions
are approximated by the exact asymptotic result  at vanishing
distance, the well known additive approximation of London dispersion forces
due to Hamaker. However, the description of retardation effects 
that is known since the time of Casimir is completely neglected for 
lack of a tractable expression.  Here we show that it is possible to describe 
surface van der Waals forces at arbitrary distances
in one single simple equation.  The result captures the  long
sought crossover from  non-retarded (London) 
to retarded (Casimir) interactions, the effect of polarization in condensed
media and the full suppression of  retarded 
interactions at large distance.  This is achieved with similar accuracy 
and the same material properties that are used to approximate the Hamaker 
constant in conventional applications. The results show that at ambient
temperature, retardation effects significantly change the power law
exponent of the conventional Hamaker result for distances of just a
few nanometers.
\end{abstract}



\maketitle



Van der Waals interactions are at the heart of chemical physics.
Yet, the standard textbook answer on their essential charcteristic
is the well known inverse sixth power dependence on the distance.
This is a largely biased statement
towards the London picture of molecular interactions, 
which treats intermolecular forces as a result of classical
electrodynamic fluctuations. At distances
of just a few nanometers, molecular interactions develop a different,
faster decay  that results from purely quantum electrodynamic fluctuations
and was  first described in the seminal work of Casimir (c.f.
\onlinecite{power01} and \onlinecite{lamoreaux07} for a hystorical
perspective).

An aclaimed unification of these two complementary views on the range
of molecular interactions has been known for a long time.\cite{dzyaloshinskii61}
The Dzyaloshinskii-Lifshitz-Pitaevskii theory (DLP) for surface forces
accross a medium  is the solution of the full thermal quantum
electrodynamic field theory for the forces between two plates across a
dielectric continuum. As a result, it   generalizes in one
single equation the Hamaker theory of additive dispersive
forces, the Debye induction potential between polar and polarizable
media, and the Keesom interactions between Boltzmann distributed
dipoles. It describes the cross--over from
purely non-retarded dispersive to retarded long--range interactions, it
reduces to the Casimir-Polder formula for the retarded force between metallic
plates in vacuum, and provides  the famous result of London for the dispersion
interaction between two atoms at short distances.

Not surprisingly,
the theory has had a profound impact on fundamental
physics,\cite{lamoreaux07}
it has motivated a large number of historical
experiments,\cite{tabor69,israelachvili72,sabisky73,bevan99,mohideen98,munday09}
and retains its theoretical influence 
in promising new studies up to date.\cite{berthoumieux10,berthoumieux18}

Unfotunately,
with some exceptions the general
theory of van der Waals interactions is largely ignored in
favor of the Hamaker picture of additive dispersive
forces.\cite{israelachvili91}
The reason is simple to infer, as the power and generality
of this approach comes at the cost of a complicated and lengthy formula
that can hardly be interpreted qualitatively, except 
in special limiting cases after lengthy manipulations and
detailed knowledge on materials 
properties.\cite{safran94,parsegian06,ninham10,butt10,french10}

Here, we show that the theory of surface van der Waals
forces can be formulated as one single simple equation
that embodies simultaneously the known low and high temperature limits,
the crossover from retarded to non retarded interactions and
the far less appreciated long range exponential suppression of 
retardation effects.  The result allows us to interpret easily the main 
qualitative features of van der Waals forces for arbitrary 
distances  and provides a convenient means to 
compute  quantitatively the results in analytic form.

The  surface free energy  between two
semi-infinite macroscopic bodies, 1 and 2, separated by a third body, $m$,
of thickness $\h$, may be written in terms of the effective
Hamaker function, $A(\h)$, as $g(\h) = -\frac{A(\h)}{12\pi \h^2}$.
In practice,  $A(\h)$  is a constant at small separations only,
and develops a complicated $\h$ dependence that is given in DLP theory as:
\begin{equation}\label{eq:1stol}
  A(\h) = \frac{3}{2} k_B T \sum_{n=0}^{\infty} \,^{\prime}
\int_{r_n}^{\infty} x 
[ R^M_{1m2}(x,n)  + R^E_{1m2}(x,n) 
] e^{-x} dx
\end{equation} 
where $R^{M}_{1m2}(x,n)=\Delta^M_{1m}(x,n)\Delta^M_{2m}(x,n)$,
$R^{E}_{1m2}(x,n)=\Delta^E_{1m}(x,n)\Delta^E_{2m}(x,n)$, while
\begin{equation}
\begin{array}{lcl}
 \Delta^M_{ij} = \frac{x_i \epsilon_j - x_j\epsilon_i}{x_i \epsilon_j + x_j\epsilon_i}
 & &
 \Delta^E_{ij} = \frac{x_i - x_j}{x_i  + x_j}
\end{array}
\end{equation}
and  $x_i^2 = x^2 + (\epsilon_i - \epsilon_m) (2  \omega_n \h/c)^2$.
In these equations, the prime next to the sum implies that the $n=0$ term
has an extra factor of $1/2$. The dielectric function,  $\epsilon_i$, 
is evaluated at the imaginary angular frequencies 
$i\omega_n$, where  $\omega_n=\omega_T n$
are integer multiples of the thermal Matsubara frequency $\omega_T= \frac{2\pi
k_BT}{\hbar}$. The magnetic permitivities have been assumed equal to unity.
The lower integration limit is $r_n=2\epsilon_m^{1/2} \omega_n \h/c$; 
while $k_B$ and $\hbar$, $c$ are the usual fundamental constants.
\Eq{1stol} provides the leading order result of the full
DLP theory, and  is the starting point of most analytical
approximations for the Hamaker 
function.\cite{tabor69,hough80,prieve88,israelachvili91,safran94,ninham10,butt10}

It is conventional to split the sum into  $n=0$ and  $n>0$ contributions,
such that $A(\h)=A_{\omega=0}+A_{\omega>0}(\h)$. For the first term $n=0$
it is possible to integrate over $x$ exactly, 
yielding the well known approximation:
\begin{equation}\label{eq:static}
  A_{\omega=0}= \frac{3}{4}  
  \frac{(\epsilon_1 -\epsilon_m)}{(\epsilon_1+\epsilon_m)}
  \frac{(\epsilon_2 -\epsilon_m)}{(\epsilon_2+\epsilon_m)}
   k_B T 
\end{equation} 
For the remaining contribution $A_{\omega>0}$, 
both $R^M_{1m2}(x,n)$ and  $R^E_{1m2}(x,n)$ remain non trivial functions
of $x$, and the integral cannot be evaluated exactly by analytical
means.  However, inspired by Parsegian's insightful
monograph,\cite{parsegian06}
we perform  the integration over $x$ using a generalized 
one-point Gausss-Laguerre quadrature
rule with weight $x e^{-x}$ in the interval $[r_n,\infty]$. This yields
(supporting information):
\begin{equation}\label{eq:aproxsum}
  A_{\omega>0}(\h) = \frac{3}{2} k_B T \sum_{n=1}^{\infty} 
 R(x^*,n) [1 + r_n] e^{-r_n} 
\end{equation} 
where we have written $R=R^M_{1m2} + R^E_{1m2}$ for short
and $R(x,n)$ is evaluated at the optimized value  
\begin{equation}\label{eq:xstar}
 x^* = \frac{2+2r_n + r_n^2}{1+r_n}
\end{equation} 

Next, we transform the sum of  \Eq{aproxsum} into an integral.
Using the Euler-McLaurin formula,  we quantify the first order 
correction to this transformation as 
$\approx\frac{3}{4}k_BT (1+ \nu_T\h) e^{-\nu_T\h}$, which is a small
fraction of the full integral in most practical situations. 
Therefore, introducing the auxiliary variable $\nu_n$, 
such that $r_n=\nu_n \h$, 
and defining the constant  $\nu_T=2\epsilon_m^{1/2}\omega_T/c$, we find:
\begin{equation}\label{eq:ahigh1}
A_{\omega>0}(\h) =  
 \frac{3 \hbar c}{8\pi}  
\int_{\nu_T}^{\infty} 
  \tilde{R}(x^*,\nu)  (1 + \nu \h)  e^{-\nu \h} d\nu
\end{equation} 
where $\tilde{R}(x^*,\nu)=\epsilon_m^{-1/2} j_m^{-1}(\nu) R(x^*,\nu)$, and
$j_m =  \left ( 1 + \frac{1}{2} \frac{d \ln \epsilon_m}{d\ln \omega} \right )$.
The factor $j_m$  is close to unity in most of the frequency interval, 
and becomes strictly equal to one for interactions between two media across 
vacuum.  

Armed with this result, we can now read-off  the
essence of the crossover behavior between retarded and non retarded
interactions. In the limit of large $\h$, the integrand is 
dominated by the exponential decay,  $\tilde{R}(x^*,\nu)$
remains essentially constant and the integral can
be approximated to $\tilde{R}(x^*,\nu_T)\int  (1 + \nu \h)  e^{-\nu \h} d\nu$.
In the opposite limit,
$\h\to 0$,  the exponential function decays very slowly
and the integrand becomes dominated by the algebraic decay of
$\tilde{R}(x^*,\nu)$ that takes place at frequencies larger than a 
characteristic
frequency of the material, $\nu_e$.  Whence,  the integral is now given by
$(1+\nu_T \h) e^{-\nu_T \h} \int  \tilde{R}(x^*,\nu) d\nu$.
This analysis  highlights the  physical origin of the crossover behavior
and illustrates the mathematical complexity of the problem.
Integrals with a crossover from algebraic to exponential decay
are non-elementary functions that
cannot be possibly expressed as a finite number  
of ordinary algebraic, exponential and logarithmic functions. 

In order to circumvent this pessimistic mathematical statement, we introduce
an auxiliary exponential function, $e^{-\nu/\nu_{\infty}}$, made to mimic the
algebraic decay of $\tilde{R}(x^*,\nu_T)$. In this
mapping, $\nu_{\infty}$ is an effective material parameter 
that dictates the range were such decay becomes effective. 
With the help of this function, we then write the trivial identity:
\begin{equation}\label{eq:trivial}
A_{\omega>0}(\h) =  
 \frac{3 \hbar c}{8\pi}  
\int_{\nu_T}^{\infty} 
   \tilde{R}(x^*,\nu) e^{\frac{\nu}{\nu_{\infty}}} 
    [ e^{-\frac{\nu}{\nu_{\infty}}} (1 + \nu \h) e^{-\nu \h} ] d\nu
\end{equation} 
The function inside the square brackets shares broadly the properties
of the exact integrand in \Eq{ahigh1}, and remains convergent for all $\h$.
Whence, 
we can use it as a reference weight function and approximate the full integral
using again a one-point generalized Gauss-Laguerre quadrature.  This leads readily to the 
expression (supporting information):
\begin{equation}\label{eq:gauss-lifshitz}
A_{\omega>0}(\h) = 
 \frac{3 \hbar c}{8\pi} 
 \tilde{R}^*_{\xi}  \, \nu_{\infty}
 \frac{(\nu_T \h + 1)(\nu_{\infty} \h + 1) + \nu_{\infty}\h}
      {(\nu_{\infty}\h+1)^2 } e^{-\nu_T \h}
\end{equation} 
where   $\tilde{R}^*_{\xi}(\h)=\tilde{R}(x^*,\nu^*)\, e^{\xi}$,
$\nu^*=\nu_T + \nu_{\infty}\xi$  and $\xi$ is an adimensional
factor given by:
\begin{equation}\label{eq:nustar}
  \xi =   \frac{(\nu_T \h + 1)(\nu_{\infty} \h + 1) + 2 \nu_{\infty}\h}
               {(\nu_{\infty} \h+1)^2(\nu_T \h + 1) + (\nu_{\infty} \h +
1)\nu_{\infty}\h}
\end{equation} 
The results of \Eq{gauss-lifshitz} and \Eq{nustar}, together with \Eq{xstar},
provide an analytical approximation that describes the qualitative behavior of
$A_{\omega>0}(\h)$ in the full range from $\h=0$ to $\h\to\infty$. 
We call this the Weighted Quadrature Approximation (WQA).  
Our analysis allows us to identify 
two different inverse length scales,
$\nu_{\infty}$ and $\nu_{T}$ which dictate the
$\h$ dependence of the Hamaker function.
In the ensuing discussion we describe the qualitative behavior
that follows from the WQA, merely by assuming that the 
wave numbers $\nu_{\infty}$ and $\nu_T$  are sufficiently separated.

\begin{itemize}
\item For $\h \ll \nu_{\infty}^{-1}$, we 
retain only the leading order term
  of \Eq{gauss-lifshitz} in the limit $\h\to 0$, we find:
\begin{equation}\label{eq:limit0}
   A_{\omega>0}(\h) = \frac{3 \hbar \omega_{\infty}}{4\pi} j_m^{-1} 
          \Delta_{1m}^M \Delta_{2m}^M e
\end{equation} 
$A_{\omega>0}$ is a constant independent of $\h$,
with
\begin{equation}
   \Delta_{jm}^M =
  \left(\frac{\epsilon_j(i\omega_{\infty}) - \epsilon_m(i\omega_{\infty})}
             {\epsilon_j(i\omega_{\infty}) + \epsilon_m(i\omega_{\infty})}
  \right)
\end{equation} 
Imposing the unknown parameter
$\omega_{\infty}=\nu_{\infty}c/2\epsilon_m^{1/2}$ such that
\Eq{limit0} recovers the known value of $A_{\omega>0}$ at $\h=0$, 
our result becomes exact in this limit by construction.
\item For $\nu_{\infty}^{-1} \ll \h \ll \nu_{T}^{-1}$
$A_{\omega>0}$ recovers the expected inverse power dependence on
$\h$:\cite{dzyaloshinskii61,israelachvili72}  
\begin{equation}\label{eq:limit1}
   A_{\omega>0}(\h) = 
  \frac{3 \hbar c}{4\pi}\frac{R^*(L)}{\epsilon_m^{1/2}(i\omega^*)}
  \frac{ 1 }{\h}
\end{equation} 
${R}^*$ is now  a complicated function of
the relative permitivities, while $\omega_{\infty}\gg\omega^*\approx
c/\h\gg\omega_T$.
\item For  $\h \gg \nu_{T}^{-1}$,  the retarded interactions become strongly suppressed due to 
the $\exp(-\nu_T \h)$ factor:
\begin{equation}\label{eq:limit3}
   A_{\omega>0}(\h) =  3 k_BT \Delta_{1m}^M\Delta_{2m}^M \, e^{-\nu_T\h}
\end{equation}
with
\begin{equation}
  \Delta_{jm}^M = \left(\frac{\sqrt{\epsilon_j(i\omega_T)} - \sqrt{\epsilon_m(i\omega_T)}}
                           {\sqrt{\epsilon_j(i\omega_T)} + \sqrt{i\epsilon_m(\omega_T)}}
                \right) 
\end{equation} 
The expression in \Eq{limit3} is the exact result  for large $\h$ at finite
temperature.\cite{ninham70} In this limit
all the dielectric functions are calculated at
$i\omega_T$ and amount barely to $\epsilon_i(i\omega_T) = n_i^2$, where $n_i$
may be identified in simple materials  with the refractive index in the
visible region.\cite{parsegian06,israelachvili91} Notice that
$A_{\omega>0}$ vanishes for $\h > \nu_T^{-1}$ and only the static 
contribution to the
van der Waals forces, \Eq{static} remains.\cite{parsegian06,butt10} Often, in analytical calculations the
low temperature limit $\nu_T\to 0$ has been considered, so that this exponential
suppression does not take place.\cite{dzyaloshinskii61,cheng88,parsegian06} At ambient temperature, however, 
this can become  a serious error for $\h$  in the micrometer range.
\end{itemize}

In the above paragraph  it has been the aim to emphasize the
crossover of $ A_{\omega>0}$ as $\h$  increases.
Therefore, only the leading order athermal contributions have been
retained. However,
by considering also next to leading terms, it is possible
to trace non-retarded contributions that operate
within the retardation dominated regime.  Particularly, for distances
$\h\nu_{\infty} \gg 1$,
$A_{\omega>0}$ features a non-retarded term of order $k_BT$
that adds up to the $A_{\omega=0}$ contribution of the full Hamaker function.
Furtermore, from the analysis it follows that the thermal contribution to the
retarded
interactions is a small fraction of order $\nu_T/\nu_{\infty}$
for $\h\nu_T \ll 1$
but grows steadily and becomes the only remaining source of retardation
for $\h\nu_T \gg 1$.  This qualifies analytically observation made
from numerical results on the significance of
thermal contribution to van der Waals
interactions.\cite{ninham70,parsegian70,chan77,parsegian06,obrecht07}

For specific applications, it is required to consider explicitly the
frequency dependence  of the material's dielectric response. In view
of the limited information that is usually available, it is customary
to describe the full dielectric function by a single classical damped
oscillator, such that:
\begin{equation}\label{eq:oscillator}
  \epsilon_{\alpha}(i\omega) = 1 + \frac{n_{\alpha}^2-1}{1+(\omega/\omega_e)^2}
\end{equation} 
where $n_{\alpha}$ is the refractive index of the material $\alpha$ in the visible, and $\omega_e$
is a characteristic absorption frequency with order of magnitude
similar to the material's ionization 
potential.\cite{hough80,parsegian06,prieve88,israelachvili91,bergstrom97}
This expression for the dielectric functions is quite convenient
because it provides well known analytical results for the
Hamaker constant ($\h=0$) under the approximation of 
\Eq{ahigh1}.\cite{hough80,parsegian06,prieve88,israelachvili91,bergstrom97} 
The corresponding results may be used to  gauge the unknown 
parameter $\nu_{\infty}$ required for quantitative validation of 
\Eq{gauss-lifshitz} and \Eq{fca}.

\begin{figure*}[t]
\begin{tabular}{cc}
\includegraphics[width=0.45\textwidth]{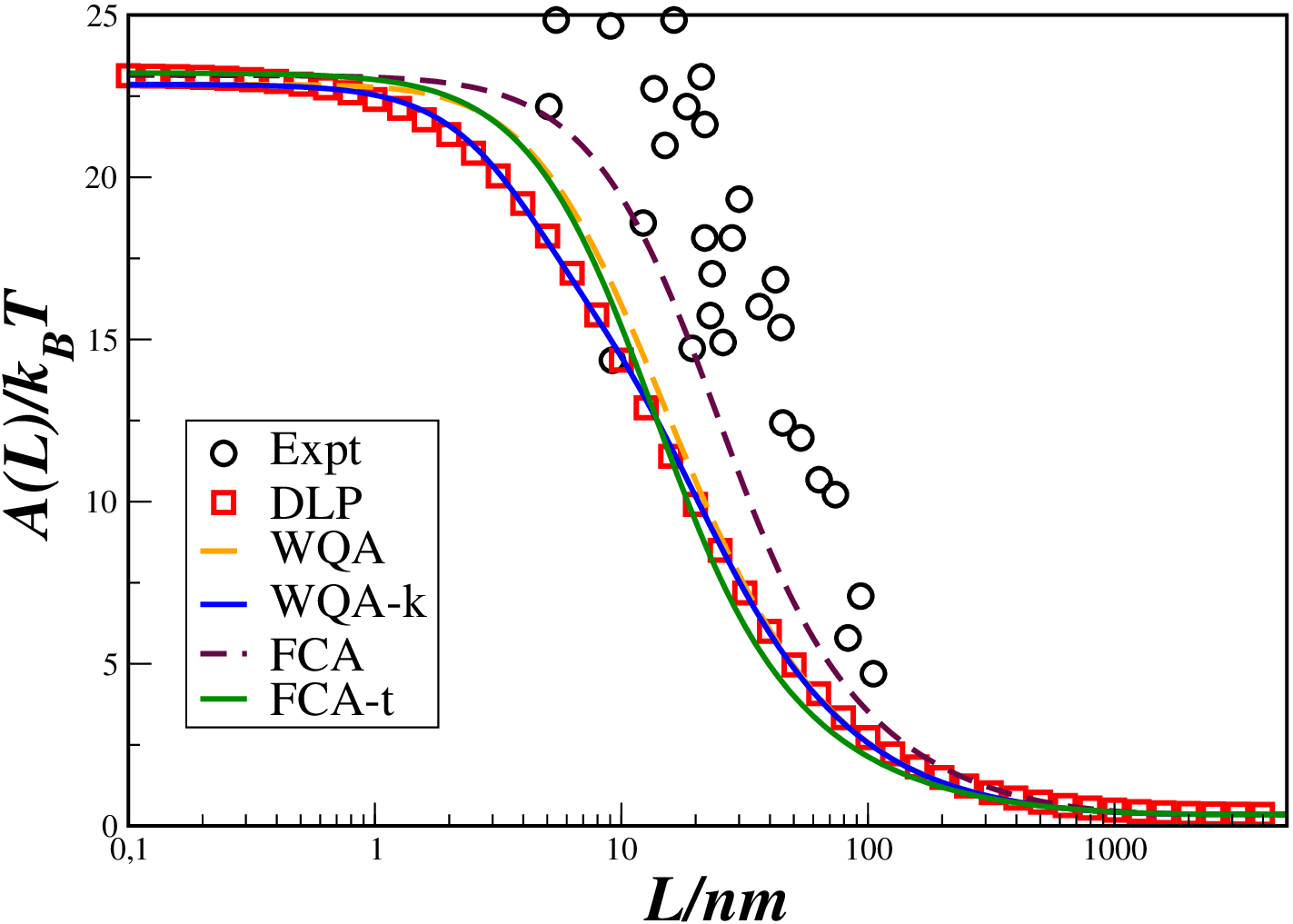} &
\includegraphics[width=0.45\textwidth]{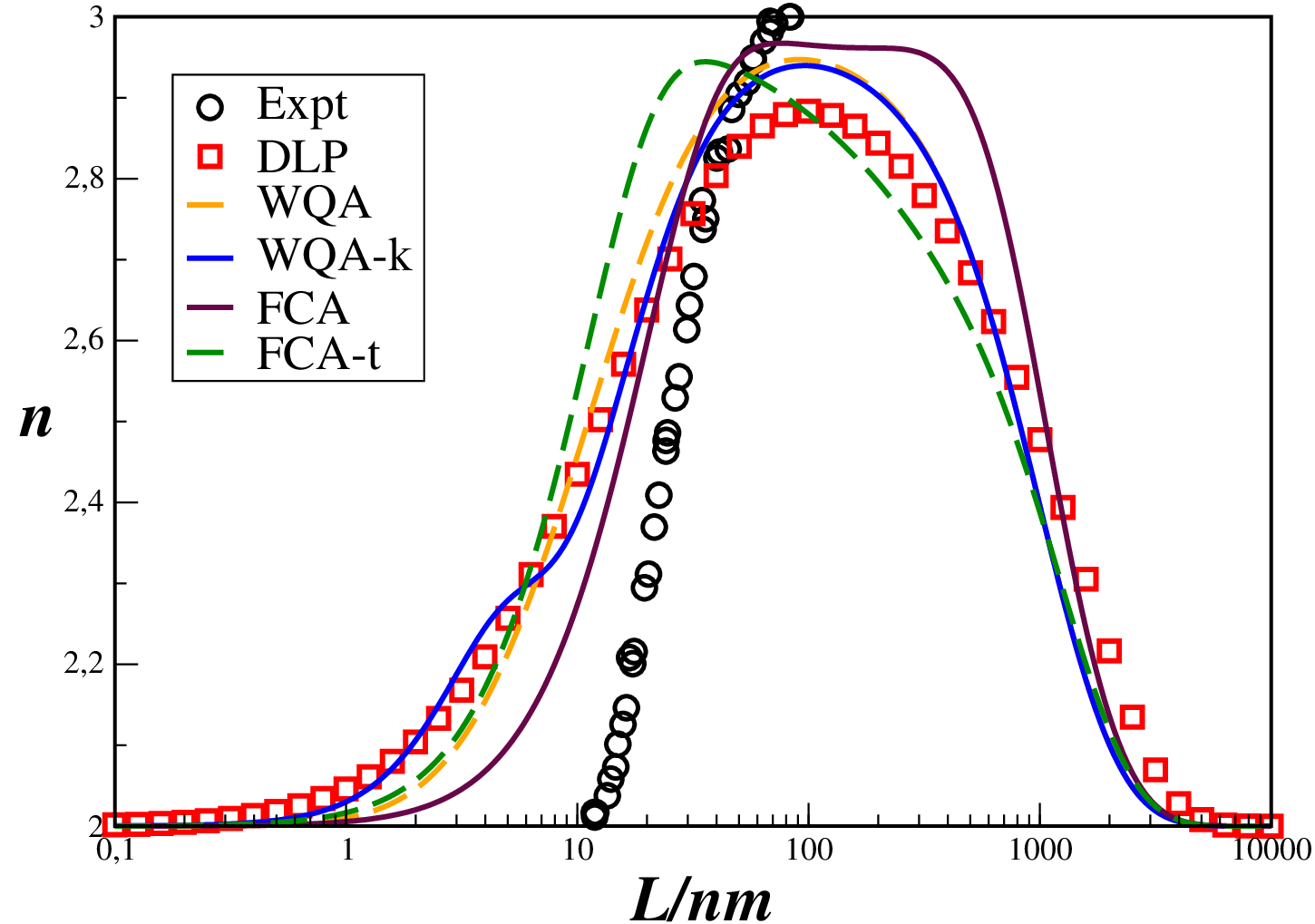} 
\end{tabular}
\caption{\label{fig:mica}
Hamaker function (left) and van der Waals effective exponent (right)
for two mica plates interacting across vacuum at T=300~K. Circles are experimental
results from Ref.\cite{israelachvili72}. Squares are exact results
from first order DLP theory \Eq{1stol}. Lines are different approximations
as described in the figure legend.
}
\end{figure*}

We use this model to describe the paradigmatic system of two mica plates
interacting across vacuum,\cite{tabor69,israelachvili72,white76,chan77} 
with the high frequency
oscillator parameters as given by 
Bergstrom.\cite{bergstrom97} Matching \Eq{gauss-lifshitz} to the 
Tabor-Winterton approximation for the Hamaker constant, readily provides
$\nu_{\infty}=1.1868\,(n_1^2+1)^{1/2}\omega_e/c$ for the cut-off frequency
(supporting information). 
This is all that is required to plot $A(\h)$ for arbitrary values of $\h$.
A comparison of the
exact first order Hamaker function obtained by integration of 
\Eq{1stol} shows very good agreement with  the WQ approximation 
(Fig.\ref{fig:mica}).
Particularly, WQA predicts the Hamaker function at large distance
almost exactly, but yields a decay rate
that is somewhat too slow.  In practice, under the approximation of  
\Eq{ahigh1} we notice that 
$\tilde{R}_{\xi}(x^*,\nu)$ is a bounded function of $\h$, with
a quadratic low order expansion  in $\h$. Therefore,
the auxiliary function $\exp(-\nu/\nu_{\infty})$ should
also exhibit a similar $\h$ decay. This can be
achieved assuming $\nu_{\infty}(\h)=(1+k(\nu_{\infty}^0\h)^2)/(1+(\nu_{\infty}^0\h)^2)\nu_{\infty}^0$,
with $k$ determined such that \Eq{gauss-lifshitz} yields the exact leading order
correction to the Hamaker function as given by \Eq{aproxsum}.
Whereas forcing this
requirement yields an exceedingly complicated algebraic expression,
we illustrate this point by assuming $k=37/20$ empirically, and
call this the WQ-k approximation (Fig.\ref{fig:mica}).
With this device,  we obtain almost
exact agreement with the Hamaker function for all $\h$. 
Notice the exact result from Lifshitz theory, as well as the models, are somewhat
off the experimental results that are accompanied in the figure for 
comparison. However, the early measurements with a prototype surface force
apparatus are presently considered as order of magnitude estimates and likely
suffer from a number of technical difficulties.\cite{chan77,white76}

Unfortunately, even with $\nu_{\infty}$ assumed constant,
the WQ approximation may be somewhat too cumbersome for some practical
applications. Fortunately, we can obtain a simpler expression for 
$A_{\omega>0}(\h)$ 
by taking into account that $\tilde{R}(x^*,\nu)$ is expected to be a 
monotonically 
decaying function in most cases (an interesting exception is the system made 
of ice/water/air). Accordingly, applying  the second mean value
theorem of definite integrals to \Eq{ahigh1}, one can write:
\begin{equation}\label{eq:2ndmvt}
A_{\omega>0}(\h) = \frac{3}{2} \frac{k_B T}{\nu_T} \tilde{R}(x^*,\nu_T)
\int_{\nu_T}^{\nu_{\infty}}
 (1 + \nu \h)  e^{-\nu \h} d\nu
\end{equation}
The above result is an exact quadrature for \Eq{ahigh1}, provided one chooses
a suitable $\h$ dependent high frequency cutoff for the
wave-number, $\nu_{\infty}$. This frequency acts effectively as a natural
ultra-violet cutoff for the integral.

Performing now a trivial integration yields:
\begin{equation}\label{eq:fca}
A_{\omega>0} =
   \frac{3 \hbar c}{8\pi\h}  
 \tilde{R}^*_{\xi=0} 
 \left [ 
(2+\frac{3}{2}\nu_T \h) e^{-\nu_T \h} - (2 + \nu_{\infty} \h ) e^{-\nu_{\infty} \h}
 \right ]
\end{equation} 
where the factor $3/2$ inside the first round parenthesis follows by
inclusion of  Euler-McLaurin corrections to lowest order.  This accounts 
for the transformation of the sum  \Eq{aproxsum} into an integral. 

For practical matters, assuming a constant value for $\nu_{\infty}$ is sufficient.  
Indeed, by matching $\nu_{\infty}$ such that $A_{\omega>0}(\h)$ yields the exact
quadrature of \Eq{ahigh1} for $\h=0$, we find:
\begin{equation}
\nu_{\infty}=t\, 
\frac{n_m \overline{n}_{1m} \overline{n}_{2m}}
{\overline{n}_{1m}+\overline{n}_{2m}} \frac{\omega_e}{c}
\end{equation} 
where $\overline{n}_{im}=(n_i^2+n_m^2)^{1/2}$ are root mean square 
indexes of refraction  and $t=\pi/\sqrt{2}$ is a  numerical 
factor (supporting information).  We call this the Frequency Cutoff approximation (FCA).

\begin{figure*}[t]
\begin{tabular}{cc}
\includegraphics[width=0.45\textwidth]{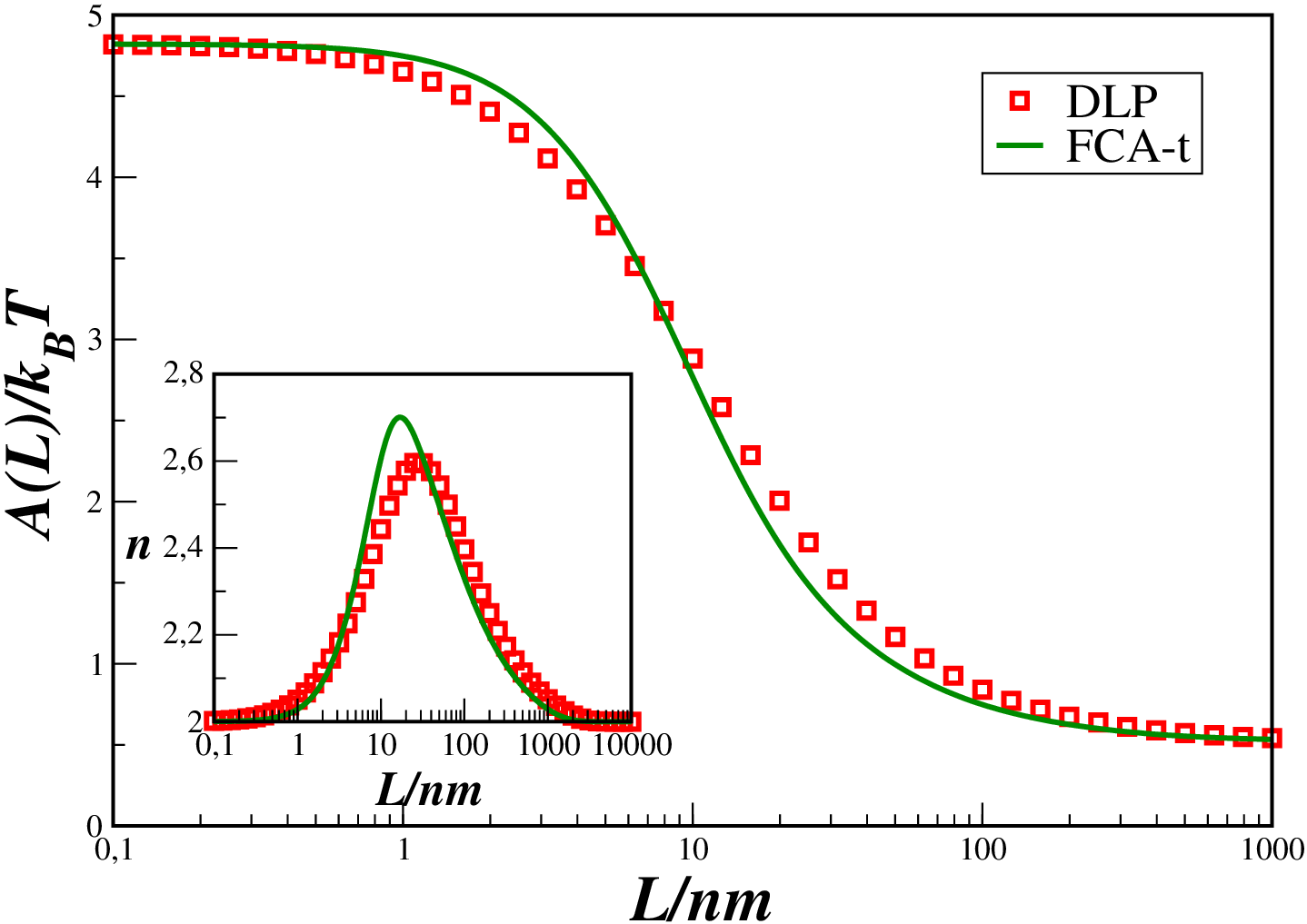} &
\includegraphics[width=0.45\textwidth]{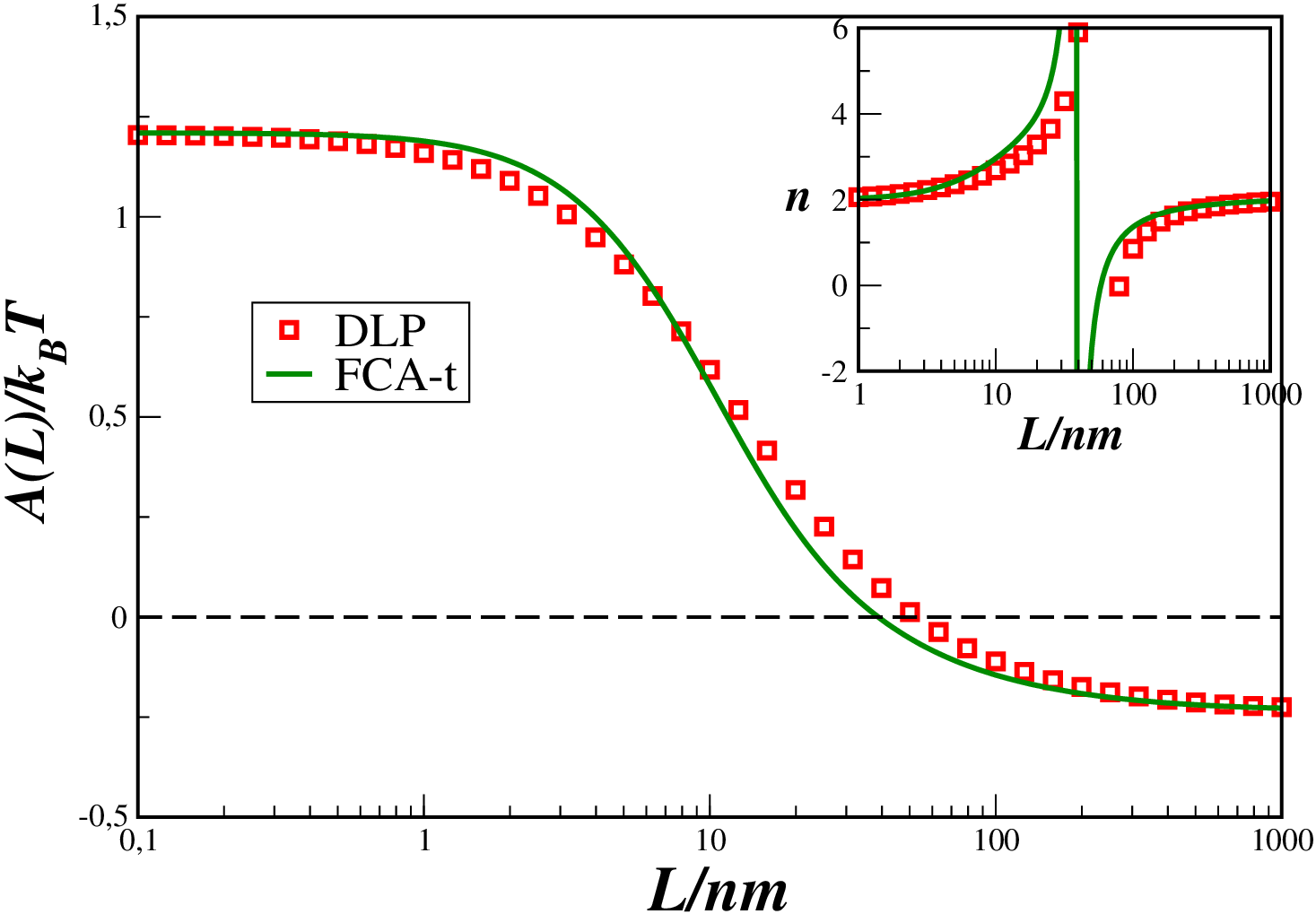} 
\end{tabular}
\caption{\label{fig:fca-t}
Comparison of Hamaker functions as obtained from DLP theory 
at T=300~K (symbols)
with results from the FCA-t approximation (lines). Inset shows
the effective van der Waals exponent. 
Left:  mica/water/mica. Right: water/octane/air (notice the sign
reversal of the Hamaker constant for this system).
}
\end{figure*}

The model is less accurate than the WQ approximation, but yields
a qualitatively good agreement in the full range of relevant distances
by using exactly the same parameters that are required to describe
the Hamaker constant in usual applications.\cite{israelachvili91} This 
contrasts with the few empirical approximations that have been previously
suggested, which only provide heuristic estimates of the first crossover 
length-scale and
completely neglect the second one.\cite{sabisky73,gregory81,cheng88}
For ultimate simplicity, we can use \Eq{fca} with the smooth function
$\tilde{R}_{\xi=0}$ replaced by a constant $\tilde{R}_{\xi=0}(\h=0)/t$,
and choose $t=4$ to improve the decay rate of $A_{\omega>0}$ (supplemental
material).
In this way, the full Hamaker function may be readily given as: 
\begin{equation}\label{eq:simple-1}
\begin{array}{ccc}
 A_{\omega > 0}(\h) & = & \displaystyle{ \frac{3\hbar c}{32\sqrt{2}\,n_m\h}
       \left ( \frac{n_1^2-n_m^2}{n_1^2+n_m^2} \frac{n_2^2-n_m^2}{n_2^2+n_m^2}
\right ) } \times \\ & & \\ & &
       \left [
 (2+\frac{3}{2}\nu_T \h) e^{-\nu_T \h} - (2 + \nu_{\infty} \h ) e^{-\nu_{\infty}
\h}
       \right ]
\end{array}
\end{equation}
Comparison of this very simple procedure (FC-t approximation) 
yields again an overall
 very good description of the full Hamaker
function, at the cost of somewhat deteriorating the large $\h$ behavior.
We stress however, that the factor $t=4$ is model independent, so
that this approach can now be applied generally for whatever system
with overall very good accuracy.
This is illustrated for interactions between two mica plates across
water and for that of octane adsorbed on water (Fig.\ref{fig:fca-t}), with
dielectric relaxation parameters taken from 
Israelachvili.\cite{israelachvili91}  Particularly, 
our approach is able to capture the sign
reversal of the Hamaker function for the water/octane/air system, 
where a description based on the Hamaker constant alone would be
unable to predict the stabilization of thick octane films
at the air/water interface.

 As a caveat, notice that in either the WQ and FC approximations,
the tractability of this approach relies on a one to one
mapping of the effective cutoff frequency, $\nu_{\infty}$ with
the resonance frequency $\omega_e$ of \Eq{oscillator}. Whence, the method
is somewhat limitted to  this simple optical description. For
a more general sum of states, either \Eq{gauss-lifshitz} or \Eq{fca} 
will still provide
a correct qualitative description by mapping $\nu_{\infty}$ to
the highest energy oscillator, provided the
optical response function  $R(x,n)$ decays  monotonously.

Aside the quantitative description discussed above for selected
systems, our approach
illustrates qualitatively a number of very relevant issues that
despite efforts (c.f. \citet{parsegian06,french10}) are
generally not recognized.  Firstly,  the
crossover from retarded to non-retarded interactions sets in for
distances of the order $c/\omega_e$. For a large number of materials,
including inorganic substrates and hydrocarbons, 
this corresponds to distances of  about
10~nm,\cite{chan77,hough80,israelachvili78,israelachvili91,parsegian06,bostrom12}
an order of magnitude less than it is often assumed.\cite{gregory81} 
Secondly, the 
rather simple crossover function $A\approx A(0)(1+\nu_{\infty}\h)^{-1}$
that is most often used in the literature yields a Hamaker constant that decays
as $\h^{-1}$ for large distances.\cite{gregory81} 
In practice, the length scales
$\nu_{\infty}$ and $\nu_T$ are not sufficiently well separated at
ambient temperature, and the decay of the van der Waals interactions
does never really attain the  $\h^{-3}$ power law expected
from the Gregory equation. This can be illustrated
by representing the effective exponent $n=d\ln g/d\ln
\h$.\cite{israelachvili72,chan77}
Clearly,
a regime of $n=3$ constant  never really sets in at ambient temperature
(Fig.\ref{fig:mica})
Rather, $n$ reaches a maximum value that is close, but smaller than the
ideal value of $3$ expected from the standard retardation regime.
This observation is not specific to a particular choice of systems
(Fig.\ref{fig:fca-t}).
It is dictated by the separation between the two constants
$\nu_{\infty}$ and $\nu_T$, which is just two orders of magnitude
at ambient temperature for most substances. 
In fact, using the FC approximation,
it is possible to show that the Hamaker function decreases to
half the Hamaker constant at distances $\h_{1/2}=4\nu_{\infty}^{-1}$,
whence, only a few times larger than the onset of retardation.
Using this estimate for $\h_{1/2}$ in either \Eq{gauss-lifshitz} or
in \Eq{fca} shows the Hamaker function obeys
a scaling form $A_{\omega>0}(\h) = A_{\omega>0}(0) u(\h/\h_{1/2})$,
with $u(x)$ a universal function, as observed
empirically by Cheng and Cole.\cite{cheng88}
As a final comment, we remark that the Gaussian 
Quadrature, which is a well known tool for numerical integration, 
can be exploited as 
an exceptional theoretical method in physics
provided one chooses suitable problem adapted weight functions.

In summary, it was shown that a simple analytical crossover functions,
\Eq{simple-1} can be used to describe the Hamaker function 
in the range from the angstrom to the micrometer using exactly
the same few material parameters that are employed 
conventionally  to calculate the Hamaker constant.  It is hoped this will
allow for a much better quantification of intermolecular forces
in a wide range of applications at the nanometric
scale.\cite{werner99,pototsky05,french10,sanchez-iglesias12,murata16}

\section{Supplementary Material}

See supplementary material for a
detailed explanation of problem adapted one
point Gauss quadrature rules, 
detailed derivation of \Eq{1stol}, \Eq{aproxsum} 
\Eq{gauss-lifshitz}, \Eq{fca}, calculation of
the high-frequency cutoff, $\nu_{\infty}$ and
parameters for the model systems.

\begin{acknowledgments}
The author would like to thank Andrew Archer and David Sibley for their
hospitality and stimulating ambience
at the Deparment of Mathematical Sciences at the University of Loughborough
where part of this work was performed. I also wish to acknowledge
funding from the `Programa Estatal de
Promoci\'on del Talento y su Empleabilidad en I+D+i', by the Spanish
Ministerio de Educaci\'on, Cultura y Deporte (Plan Estatal de Investigaci\'on
Cient\'{\i}fica y T\'ecnica y de Innovaci\'on 2013-2016)  for the generous
funding of this visit and 
to the Agencia Estatal de Investigaci\'on y Fondo Europeo de Desarrollo Regional
(FEDER) for funding under research grant FIS2017-89361-C3-2-P.
\end{acknowledgments}



\def\bibsource{/home/luis/Ciencia/tex/Patrones}
\bibliography{\bibsource/referenc}

\begin{thebibliography}{33}%
\makeatletter
\providecommand \@ifxundefined [1]{%
 \@ifx{#1\undefined}
}%
\providecommand \@ifnum [1]{%
 \ifnum #1\expandafter \@firstoftwo
 \else \expandafter \@secondoftwo
 \fi
}%
\providecommand \@ifx [1]{%
 \ifx #1\expandafter \@firstoftwo
 \else \expandafter \@secondoftwo
 \fi
}%
\providecommand \natexlab [1]{#1}%
\providecommand \enquote  [1]{``#1''}%
\providecommand \bibnamefont  [1]{#1}%
\providecommand \bibfnamefont [1]{#1}%
\providecommand \citenamefont [1]{#1}%
\providecommand \href@noop [0]{\@secondoftwo}%
\providecommand \href [0]{\begingroup \@sanitize@url \@href}%
\providecommand \@href[1]{\@@startlink{#1}\@@href}%
\providecommand \@@href[1]{\endgroup#1\@@endlink}%
\providecommand \@sanitize@url [0]{\catcode `\\12\catcode `\$12\catcode
  `\&12\catcode `\#12\catcode `\^12\catcode `\_12\catcode `\%12\relax}%
\providecommand \@@startlink[1]{}%
\providecommand \@@endlink[0]{}%
\providecommand \url  [0]{\begingroup\@sanitize@url \@url }%
\providecommand \@url [1]{\endgroup\@href {#1}{\urlprefix }}%
\providecommand \urlprefix  [0]{URL }%
\providecommand \Eprint [0]{\href }%
\providecommand \doibase [0]{http://dx.doi.org/}%
\providecommand \selectlanguage [0]{\@gobble}%
\providecommand \bibinfo  [0]{\@secondoftwo}%
\providecommand \bibfield  [0]{\@secondoftwo}%
\providecommand \translation [1]{[#1]}%
\providecommand \BibitemOpen [0]{}%
\providecommand \bibitemStop [0]{}%
\providecommand \bibitemNoStop [0]{.\EOS\space}%
\providecommand \EOS [0]{\spacefactor3000\relax}%
\providecommand \BibitemShut  [1]{\csname bibitem#1\endcsname}%
\let\auto@bib@innerbib\@empty
\bibitem [{\citenamefont {Power}(2001)}]{power01}%
  \BibitemOpen
  \bibfield  {author} {\bibinfo {author} {\bibfnamefont {E.~A.}\ \bibnamefont
  {Power}},\ }\href {\doibase 10.1088/0143-0807/22/4/322} {\bibfield  {journal}
  {\bibinfo  {journal} {Eur. J. Phys.}\ }\textbf {\bibinfo {volume} {22}},\
  \bibinfo {pages} {453} (\bibinfo {year} {2001})}\BibitemShut {NoStop}%
\bibitem [{\citenamefont {Lamoreaux}(2007)}]{lamoreaux07}%
  \BibitemOpen
  \bibfield  {author} {\bibinfo {author} {\bibfnamefont {S.~K.}\ \bibnamefont
  {Lamoreaux}},\ }\href@noop {} {\bibfield  {journal} {\bibinfo  {journal}
  {Phys. Today}\ }\textbf {\bibinfo {volume} {60}},\ \bibinfo {pages} {40}
  (\bibinfo {year} {2007})}\BibitemShut {NoStop}%
\bibitem [{\citenamefont {Dzyaloshinskii}\ \emph {et~al.}(1961)\citenamefont
  {Dzyaloshinskii}, \citenamefont {Lifshitz},\ and\ \citenamefont
  {Pitaevskii}}]{dzyaloshinskii61}%
  \BibitemOpen
  \bibfield  {author} {\bibinfo {author} {\bibfnamefont {I.~E.}\ \bibnamefont
  {Dzyaloshinskii}}, \bibinfo {author} {\bibfnamefont {E.~M.}\ \bibnamefont
  {Lifshitz}}, \ and\ \bibinfo {author} {\bibfnamefont {L.~P.}\ \bibnamefont
  {Pitaevskii}},\ }
  {\bibfield  {journal} {\bibinfo  {journal} {Soviet Physics Uspekhi}\ }\textbf
  {\bibinfo {volume} {4}},\ \bibinfo {pages} {153} (\bibinfo {year}
  {1961})}\BibitemShut {NoStop}%
\bibitem [{\citenamefont {Tabor}\ and\ \citenamefont
  {Winterton}(1969)}]{tabor69}%
  \BibitemOpen
  \bibfield  {author} {\bibinfo {author} {\bibfnamefont {D.}~\bibnamefont
  {Tabor}}\ and\ \bibinfo {author} {\bibfnamefont {R.~H.~S.}\ \bibnamefont
  {Winterton}},\ }\href {\doibase 10.1098/rspa.1969.0169} {\bibfield  {journal}
  {\bibinfo  {journal} {Proc. R. Soc. Lond. A}\ }\textbf {\bibinfo {volume}
  {312}},\ \bibinfo {pages} {435} (\bibinfo {year} {1969})},\ 
  \BibitemShut {NoStop}%
\bibitem [{\citenamefont {Israelachvili}\ and\ \citenamefont
  {Tabor}(1972)}]{israelachvili72}%
  \BibitemOpen
  \bibfield  {author} {\bibinfo {author} {\bibfnamefont {J.~N.}\ \bibnamefont
  {Israelachvili}}\ and\ \bibinfo {author} {\bibfnamefont {D.}~\bibnamefont
  {Tabor}},\ }\href {\doibase 10.1098/rspa.1972.0162} {\bibfield  {journal}
  {\bibinfo  {journal} {Proceedings of the Royal Society of London A:
  Mathematical, Physical and Engineering Sciences}\ }\textbf {\bibinfo {volume}
  {331}},\ \bibinfo {pages} {19} (\bibinfo {year} {1972})},\ 
  \BibitemShut {NoStop}%
\bibitem [{\citenamefont {Sabisky}\ and\ \citenamefont
  {Anderson}(1973)}]{sabisky73}%
  \BibitemOpen
  \bibfield  {author} {\bibinfo {author} {\bibfnamefont {E.~S.}\ \bibnamefont
  {Sabisky}}\ and\ \bibinfo {author} {\bibfnamefont {C.~H.}\ \bibnamefont
  {Anderson}},\ }\href {\doibase 10.1103/PhysRevA.7.790} {\bibfield  {journal}
  {\bibinfo  {journal} {Phys. Rev. A}\ }\textbf {\bibinfo {volume} {7}},\
  \bibinfo {pages} {790} (\bibinfo {year} {1973})}\BibitemShut {NoStop}%
\bibitem [{\citenamefont {Bevan}\ and\ \citenamefont {Prieve}(1999)}]{bevan99}%
  \BibitemOpen
  \bibfield  {author} {\bibinfo {author} {\bibfnamefont {M.~A.}\ \bibnamefont
  {Bevan}}\ and\ \bibinfo {author} {\bibfnamefont {D.~C.}\ \bibnamefont
  {Prieve}},\ }\href {\doibase 10.1021/la981381l} {\bibfield  {journal}
  {\bibinfo  {journal} {Langmuir}\ }\textbf {\bibinfo {volume} {15}},\ \bibinfo
  {pages} {7925} (\bibinfo {year} {1999})},\ 
  \BibitemShut {NoStop}%
\bibitem [{\citenamefont {Mohideen}\ and\ \citenamefont
  {Roy}(1998)}]{mohideen98}%
  \BibitemOpen
  \bibfield  {author} {\bibinfo {author} {\bibfnamefont {U.}~\bibnamefont
  {Mohideen}}\ and\ \bibinfo {author} {\bibfnamefont {A.}~\bibnamefont {Roy}},\
  }\href {\doibase 10.1103/PhysRevLett.81.4549} {\bibfield  {journal} {\bibinfo
   {journal} {Phys. Rev. Lett.}\ }\textbf {\bibinfo {volume} {81}},\ \bibinfo
  {pages} {4549} (\bibinfo {year} {1998})}\BibitemShut {NoStop}%
\bibitem [{\citenamefont {Munday}\ \emph {et~al.}(2009)\citenamefont {Munday},
  \citenamefont {Capasso},\ and\ \citenamefont {Parsegian}}]{munday09}%
  \BibitemOpen
  \bibfield  {author} {\bibinfo {author} {\bibfnamefont {J.~N.}\ \bibnamefont
  {Munday}}, \bibinfo {author} {\bibfnamefont {F.}~\bibnamefont {Capasso}}, \
  and\ \bibinfo {author} {\bibfnamefont {V.~A.}\ \bibnamefont {Parsegian}},\
  }\href@noop {} {\bibfield  {journal} {\bibinfo  {journal} {Nature}\ }\textbf
  {\bibinfo {volume} {457}},\ \bibinfo {pages} {170} (\bibinfo {year}
  {2009})}\BibitemShut {NoStop}%
\bibitem [{\citenamefont {Berthoumieux}\ and\ \citenamefont
  {Maggs}(2010)}]{berthoumieux10}%
  \BibitemOpen
  \bibfield  {author} {\bibinfo {author} {\bibfnamefont {H.}~\bibnamefont
  {Berthoumieux}}\ and\ \bibinfo {author} {\bibfnamefont {A.~C.}\ \bibnamefont
  {Maggs}},\ }\href@noop {} {\bibfield  {journal} {\bibinfo  {journal}
  {Europhys. Lett}\ }\textbf {\bibinfo {volume} {91}},\ \bibinfo {pages}
  {56006} (\bibinfo {year} {2010})}\BibitemShut {NoStop}%
\bibitem [{\citenamefont {Berthoumieux}(2018)}]{berthoumieux18}%
  \BibitemOpen
  \bibfield  {author} {\bibinfo {author} {\bibfnamefont {H.}~\bibnamefont
  {Berthoumieux}},\ }\href {\doibase 10.1063/1.5012828} {\bibfield  {journal}
  {\bibinfo  {journal} {J. Chem. Phys.}\ }\textbf {\bibinfo {volume} {148}},\
  \bibinfo {pages} {104504} (\bibinfo {year} {2018})},\ 
  \BibitemShut {NoStop}%
\bibitem [{\citenamefont {Israelachvili}(1991)}]{israelachvili91}%
  \BibitemOpen
  \bibfield  {author} {\bibinfo {author} {\bibfnamefont {J.~N.}\ \bibnamefont
  {Israelachvili}},\ }\href@noop {} {\emph {\bibinfo {title} {Intermolecular
  and Surfaces Forces}}},\ \bibinfo {edition} {2nd}\ ed.\ (\bibinfo
  {publisher} {Academic Press},\ \bibinfo {address} {London},\ \bibinfo {year}
  {1991})\BibitemShut {NoStop}%
\bibitem [{\citenamefont {Safran}(1994)}]{safran94}%
  \BibitemOpen
  \bibfield  {author} {\bibinfo {author} {\bibfnamefont {S.~A.}\ \bibnamefont
  {Safran}},\ }\href@noop {} {\emph {\bibinfo {title} {Statistical
  Thermodynamics of Surfaces, Interfaces and Membranes}}},\ \bibinfo {edition}
  {1st}\ ed.\ (\bibinfo  {publisher} {Addison-Wesley},\ \bibinfo {address}
  {Reading},\ \bibinfo {year} {1994})\BibitemShut {NoStop}%
\bibitem [{\citenamefont {Parsegian}(2006)}]{parsegian06}%
  \BibitemOpen
  \bibfield  {author} {\bibinfo {author} {\bibfnamefont {V.~A.}\ \bibnamefont
  {Parsegian}},\ }\href@noop {} {\emph {\bibinfo {title} {Van der Waals
  Forces}}}\ (\bibinfo  {publisher} {Cambridge University Press},\ \bibinfo
  {address} {Cambridge},\ \bibinfo {year} {2006})\ pp.\ \bibinfo {pages}
  {1--311}\BibitemShut {NoStop}%
\bibitem [{\citenamefont {Ninham}\ and\ \citenamefont
  {Nostro}(2010)}]{ninham10}%
  \BibitemOpen
  \bibfield  {author} {\bibinfo {author} {\bibfnamefont {B.~W.}\ \bibnamefont
  {Ninham}}\ and\ \bibinfo {author} {\bibfnamefont {P.~L.}\ \bibnamefont
  {Nostro}},\ }\href@noop {} {\emph {\bibinfo {title} {Molecular Forces and
  Self Assembly: In Colloids, Nanoscience and Biology}}}\ (\bibinfo
  {publisher} {Cambridge University Press},\ \bibinfo {address} {Cambridge},\
  \bibinfo {year} {2010})\BibitemShut {NoStop}%
\bibitem [{\citenamefont {Butt}\ and\ \citenamefont {Kappl}(2010)}]{butt10}%
  \BibitemOpen
  \bibfield  {author} {\bibinfo {author} {\bibfnamefont {H.-J.}\ \bibnamefont
  {Butt}}\ and\ \bibinfo {author} {\bibfnamefont {M.}~\bibnamefont {Kappl}},\
  }\href@noop {} {\emph {\bibinfo {title} {Surface and Interfacial Forces}}}\
  (\bibinfo  {publisher} {Wiley-VCH},\ \bibinfo {address} {Weinheim},\ \bibinfo
  {year} {2010})\BibitemShut {NoStop}%
\bibitem [{\citenamefont {French}\ \emph {et~al.}(2010)\citenamefont {French},
  \citenamefont {Parsegian}, \citenamefont {Podgornik}, \citenamefont {Rajter},
  \citenamefont {Jagota}, \citenamefont {Luo}, \citenamefont {Asthagiri},
  \citenamefont {Chaudhury}, \citenamefont {Chiang}, \citenamefont {Granick},
  \citenamefont {Kalinin}, \citenamefont {Kardar}, \citenamefont {Kjellander},
  \citenamefont {Langreth}, \citenamefont {Lewis}, \citenamefont {Lustig},
  \citenamefont {Wesolowski}, \citenamefont {Wettlaufer}, \citenamefont
  {Ching}, \citenamefont {Finnis}, \citenamefont {Houlihan}, \citenamefont {von
  Lilienfeld}, \citenamefont {van Oss},\ and\ \citenamefont {Zemb}}]{french10}%
  \BibitemOpen
  \bibfield  {author} {\bibinfo {author} {\bibfnamefont {R.~H.}\ \bibnamefont
  {French}}, \bibinfo {author} {\bibfnamefont {V.~A.}\ \bibnamefont
  {Parsegian}}, \bibinfo {author} {\bibfnamefont {R.}~\bibnamefont
  {Podgornik}}, \bibinfo {author} {\bibfnamefont {R.~F.}\ \bibnamefont
  {Rajter}}, \bibinfo {author} {\bibfnamefont {A.}~\bibnamefont {Jagota}},
  \bibinfo {author} {\bibfnamefont {J.}~\bibnamefont {Luo}}, \bibinfo {author}
  {\bibfnamefont {D.}~\bibnamefont {Asthagiri}}, \bibinfo {author}
  {\bibfnamefont {M.~K.}\ \bibnamefont {Chaudhury}}, \bibinfo {author}
  {\bibfnamefont {Y.-m.}\ \bibnamefont {Chiang}}, \bibinfo {author}
  {\bibfnamefont {S.}~\bibnamefont {Granick}}, \bibinfo {author} {\bibfnamefont
  {S.}~\bibnamefont {Kalinin}}, \bibinfo {author} {\bibfnamefont
  {M.}~\bibnamefont {Kardar}}, \bibinfo {author} {\bibfnamefont
  {R.}~\bibnamefont {Kjellander}}, \bibinfo {author} {\bibfnamefont {D.~C.}\
  \bibnamefont {Langreth}}, \bibinfo {author} {\bibfnamefont {J.}~\bibnamefont
  {Lewis}}, \bibinfo {author} {\bibfnamefont {S.}~\bibnamefont {Lustig}},
  \bibinfo {author} {\bibfnamefont {D.}~\bibnamefont {Wesolowski}}, \bibinfo
  {author} {\bibfnamefont {J.~S.}\ \bibnamefont {Wettlaufer}}, \bibinfo
  {author} {\bibfnamefont {W.-Y.}\ \bibnamefont {Ching}}, \bibinfo {author}
  {\bibfnamefont {M.}~\bibnamefont {Finnis}}, \bibinfo {author} {\bibfnamefont
  {F.}~\bibnamefont {Houlihan}}, \bibinfo {author} {\bibfnamefont {O.~A.}\
  \bibnamefont {von Lilienfeld}}, \bibinfo {author} {\bibfnamefont {C.~J.}\
  \bibnamefont {van Oss}}, \ and\ \bibinfo {author} {\bibfnamefont
  {T.}~\bibnamefont {Zemb}},\ }\href {\doibase 10.1103/RevModPhys.82.1887}
  {\bibfield  {journal} {\bibinfo  {journal} {Rev. Mod. Phys.}\ }\textbf
  {\bibinfo {volume} {82}},\ \bibinfo {pages} {1887} (\bibinfo {year}
  {2010})}\BibitemShut {NoStop}%
\bibitem [{\citenamefont {Hough}\ and\ \citenamefont {White}(1980)}]{hough80}%
  \BibitemOpen
  \bibfield  {author} {\bibinfo {author} {\bibfnamefont {D.~B.}\ \bibnamefont
  {Hough}}\ and\ \bibinfo {author} {\bibfnamefont {L.~R.}\ \bibnamefont
  {White}},\ }\href@noop {} {\bibfield  {journal} {\bibinfo  {journal} {Adv.
  Colloid Interface Sci.}\ }\textbf {\bibinfo {volume} {14}},\ \bibinfo {pages}
  {3} (\bibinfo {year} {1980})}\BibitemShut {NoStop}%
\bibitem [{\citenamefont {Prieve}\ and\ \citenamefont
  {Russel}(1988)}]{prieve88}%
  \BibitemOpen
  \bibfield  {author} {\bibinfo {author} {\bibfnamefont {D.~C.}\ \bibnamefont
  {Prieve}}\ and\ \bibinfo {author} {\bibfnamefont {W.~B.}\ \bibnamefont
  {Russel}},\ }
  {\bibfield  {journal} {\bibinfo  {journal} {J. Colloid. Interface Sci.}\
  }\textbf {\bibinfo {volume} {125}},\ \bibinfo {pages} {1} (\bibinfo {year}
  {1988})}\BibitemShut {NoStop}%
\bibitem [{\citenamefont {Ninham}\ and\ \citenamefont
  {Parsegian}(1970)}]{ninham70}%
  \BibitemOpen
  \bibfield  {author} {\bibinfo {author} {\bibfnamefont {B.~W.}\ \bibnamefont
  {Ninham}}\ and\ \bibinfo {author} {\bibfnamefont {V.~A.}\ \bibnamefont
  {Parsegian}},\ }\href@noop {} {\bibfield  {journal} {\bibinfo  {journal}
  {Biophys. J.}\ }\textbf {\bibinfo {volume} {10}},\ \bibinfo {pages} {646}
  (\bibinfo {year} {1970})}\BibitemShut {NoStop}%
\bibitem [{\citenamefont {Cheng}\ and\ \citenamefont {Cole}(1988)}]{cheng88}%
  \BibitemOpen
  \bibfield  {author} {\bibinfo {author} {\bibfnamefont {E.}~\bibnamefont
  {Cheng}}\ and\ \bibinfo {author} {\bibfnamefont {M.~W.}\ \bibnamefont
  {Cole}},\ }\href {\doibase 10.1103/PhysRevB.38.987} {\bibfield  {journal}
  {\bibinfo  {journal} {Phys. Rev. B}\ }\textbf {\bibinfo {volume} {38}},\
  \bibinfo {pages} {987} (\bibinfo {year} {1988})}\BibitemShut {NoStop}%
\bibitem [{\citenamefont {Parsegian}\ and\ \citenamefont
  {Ninham}(1970)}]{parsegian70}%
  \BibitemOpen
  \bibfield  {author} {\bibinfo {author} {\bibfnamefont {V.~A.}\ \bibnamefont
  {Parsegian}}\ and\ \bibinfo {author} {\bibfnamefont {B.~W.}\ \bibnamefont
  {Ninham}},\ }\href@noop {} {\bibfield  {journal} {\bibinfo  {journal}
  {Biophys. J.}\ }\textbf {\bibinfo {volume} {10}},\ \bibinfo {pages} {664}
  (\bibinfo {year} {1970})}\BibitemShut {NoStop}%
\bibitem [{\citenamefont {Chan}\ and\ \citenamefont {Richmond}(1977)}]{chan77}%
  \BibitemOpen
  \bibfield  {author} {\bibinfo {author} {\bibfnamefont {D.}~\bibnamefont
  {Chan}}\ and\ \bibinfo {author} {\bibfnamefont {P.}~\bibnamefont
  {Richmond}},\ }\href {\doibase 10.1098/rspa.1977.0027} {\bibfield  {journal}
  {\bibinfo  {journal} {Proc. R. Soc. Lond. A}\ }\textbf {\bibinfo {volume}
  {353}},\ \bibinfo {pages} {163} (\bibinfo {year} {1977})},\ 
  \BibitemShut {NoStop}%
\bibitem [{\citenamefont {Obrecht}\ \emph {et~al.}(2007)\citenamefont
  {Obrecht}, \citenamefont {Wild}, \citenamefont {Antezza}, \citenamefont
  {Pitaevskii}, \citenamefont {Stringari},\ and\ \citenamefont
  {Cornell}}]{obrecht07}%
  \BibitemOpen
  \bibfield  {author} {\bibinfo {author} {\bibfnamefont {J.~M.}\ \bibnamefont
  {Obrecht}}, \bibinfo {author} {\bibfnamefont {R.~J.}\ \bibnamefont {Wild}},
  \bibinfo {author} {\bibfnamefont {M.}~\bibnamefont {Antezza}}, \bibinfo
  {author} {\bibfnamefont {L.~P.}\ \bibnamefont {Pitaevskii}}, \bibinfo
  {author} {\bibfnamefont {S.}~\bibnamefont {Stringari}}, \ and\ \bibinfo
  {author} {\bibfnamefont {E.~A.}\ \bibnamefont {Cornell}},\ }\href {\doibase
  10.1103/PhysRevLett.98.063201} {\bibfield  {journal} {\bibinfo  {journal}
  {Phys. Rev. Lett.}\ }\textbf {\bibinfo {volume} {98}},\ \bibinfo {pages}
  {063201} (\bibinfo {year} {2007})}\BibitemShut {NoStop}%
\bibitem [{\citenamefont {Bergstr{\"o}m}(1997)}]{bergstrom97}%
  \BibitemOpen
  \bibfield  {author} {\bibinfo {author} {\bibfnamefont {L.}~\bibnamefont
  {Bergstr{\"o}m}},\ } {\bibfield  {journal}
  {\bibinfo  {journal} {Adv. Colloid Interface Sci.}\ }\textbf {\bibinfo
  {volume} {70}},\ \bibinfo {pages} {125 } (\bibinfo {year}
  {1997})}\BibitemShut {NoStop}%
\bibitem [{\citenamefont {White}\ \emph {et~al.}(1976)\citenamefont {White},
  \citenamefont {Israelachvili},\ and\ \citenamefont {Ninham}}]{white76}%
  \BibitemOpen
  \bibfield  {author} {\bibinfo {author} {\bibfnamefont {L.~R.}\ \bibnamefont
  {White}}, \bibinfo {author} {\bibfnamefont {J.~N.}\ \bibnamefont
  {Israelachvili}}, \ and\ \bibinfo {author} {\bibfnamefont {B.~W.}\
  \bibnamefont {Ninham}},\ }\href {\doibase 10.1039/F19767202526} {\bibfield
  {journal} {\bibinfo  {journal} {J. Chem. Soc.{,} Faraday Trans. 1}\ }\textbf
  {\bibinfo {volume} {72}},\ \bibinfo {pages} {2526} (\bibinfo {year}
  {1976})}\BibitemShut {NoStop}%
\bibitem [{\citenamefont {Gregory}(1981)}]{gregory81}%
  \BibitemOpen
  \bibfield  {author} {\bibinfo {author} {\bibfnamefont {J.}~\bibnamefont
  {Gregory}},\ } {\bibfield  {journal} {\bibinfo  {journal} {J. Colloid. Interface Sci.}\ }\textbf {\bibinfo {volume} {83}},\ \bibinfo {pages} {138 } (\bibinfo {year} {1981})}\BibitemShut {NoStop}%
\bibitem [{\citenamefont {Israelachvili}\ and\ \citenamefont
  {Adams}(1978)}]{israelachvili78}%
  \BibitemOpen
  \bibfield  {author} {\bibinfo {author} {\bibfnamefont {J.~N.}\ \bibnamefont
  {Israelachvili}}\ and\ \bibinfo {author} {\bibfnamefont {G.~E.}\ \bibnamefont
  {Adams}},\ }\href {\doibase 10.1039/F19787400975} {\bibfield  {journal}
  {\bibinfo  {journal} {J. Chem. Soc.{,} Faraday Trans. 1}\ }\textbf {\bibinfo
  {volume} {74}},\ \bibinfo {pages} {975} (\bibinfo {year} {1978})}\BibitemShut
  {NoStop}%
\bibitem [{\citenamefont {Bostr\"om}\ \emph {et~al.}(2012)\citenamefont
  {Bostr\"om}, \citenamefont {Sernelius}, \citenamefont {Brevik},\ and\
  \citenamefont {Ninham}}]{bostrom12}%
  \BibitemOpen
  \bibfield  {author} {\bibinfo {author} {\bibfnamefont {M.}~\bibnamefont
  {Bostr\"om}}, \bibinfo {author} {\bibfnamefont {B.~E.}\ \bibnamefont
  {Sernelius}}, \bibinfo {author} {\bibfnamefont {I.}~\bibnamefont {Brevik}}, \
  and\ \bibinfo {author} {\bibfnamefont {B.~W.}\ \bibnamefont {Ninham}},\
  }\href {\doibase 10.1103/PhysRevA.85.010701} {\bibfield  {journal} {\bibinfo
  {journal} {Phys. Rev. A}\ }\textbf {\bibinfo {volume} {85}},\ \bibinfo
  {pages} {010701} (\bibinfo {year} {2012})}\BibitemShut {NoStop}%
\bibitem [{\citenamefont {Werner}\ \emph {et~al.}(1999)\citenamefont {Werner},
  \citenamefont {Muller}, \citenamefont {Schmid},\ and\ \citenamefont
  {Binder}}]{werner99}%
  \BibitemOpen
  \bibfield  {author} {\bibinfo {author} {\bibfnamefont {A.}~\bibnamefont
  {Werner}}, \bibinfo {author} {\bibfnamefont {M.}~\bibnamefont {Muller}},
  \bibinfo {author} {\bibfnamefont {F.}~\bibnamefont {Schmid}}, \ and\ \bibinfo
  {author} {\bibfnamefont {K.}~\bibnamefont {Binder}},\ }\href {\doibase
  10.1063/1.478164} {\bibfield  {journal} {\bibinfo  {journal} {J. Chem.
  Phys.}\ }\textbf {\bibinfo {volume} {110}},\ \bibinfo {pages} {1221}
  (\bibinfo {year} {1999})}\BibitemShut {NoStop}%
\bibitem [{\citenamefont {Pototsky}\ \emph {et~al.}(2005)\citenamefont
  {Pototsky}, \citenamefont {Bestehorn}, \citenamefont {Merkt},\ and\
  \citenamefont {Thiele}}]{pototsky05}%
  \BibitemOpen
  \bibfield  {author} {\bibinfo {author} {\bibfnamefont {A.}~\bibnamefont
  {Pototsky}}, \bibinfo {author} {\bibfnamefont {M.}~\bibnamefont {Bestehorn}},
  \bibinfo {author} {\bibfnamefont {D.}~\bibnamefont {Merkt}}, \ and\ \bibinfo
  {author} {\bibfnamefont {U.}~\bibnamefont {Thiele}},\ }\href {\doibase
  10.1063/1.1927512} {\bibfield  {journal} {\bibinfo  {journal} {J. Chem.
  Phys.}\ }\textbf {\bibinfo {volume} {122}},\ \bibinfo {pages} {224711}
  (\bibinfo {year} {2005})},\ \BibitemShut {NoStop}%
\bibitem [{\citenamefont {S{\'a}nchez-Iglesias}\ \emph
  {et~al.}(2012)\citenamefont {S{\'a}nchez-Iglesias}, \citenamefont
  {Grzelczak}, \citenamefont {Altantzis}, \citenamefont {Goris}, \citenamefont
  {P{\'e}rez-Juste}, \citenamefont {Bals}, \citenamefont {{Van Tendeloo}},
  \citenamefont {Donaldson}, \citenamefont {Chmelka}, \citenamefont
  {Israelachvili},\ and\ \citenamefont {Liz-Marz{\'a}n}}]{sanchez-iglesias12}%
  \BibitemOpen
  \bibfield  {author} {\bibinfo {author} {\bibfnamefont {A.}~\bibnamefont
  {S{\'a}nchez-Iglesias}}, \bibinfo {author} {\bibfnamefont {M.}~\bibnamefont
  {Grzelczak}}, \bibinfo {author} {\bibfnamefont {T.}~\bibnamefont
  {Altantzis}}, \bibinfo {author} {\bibfnamefont {B.}~\bibnamefont {Goris}},
  \bibinfo {author} {\bibfnamefont {J.}~\bibnamefont {P{\'e}rez-Juste}},
  \bibinfo {author} {\bibfnamefont {S.}~\bibnamefont {Bals}}, \bibinfo {author}
  {\bibfnamefont {G.}~\bibnamefont {{Van Tendeloo}}}, \bibinfo {author}
  {\bibfnamefont {S.~H.}\ \bibnamefont {Donaldson}}, \bibinfo {author}
  {\bibfnamefont {B.~F.}\ \bibnamefont {Chmelka}}, \bibinfo {author}
  {\bibfnamefont {J.~N.}\ \bibnamefont {Israelachvili}}, \ and\ \bibinfo
  {author} {\bibfnamefont {L.~M.}\ \bibnamefont {Liz-Marz{\'a}n}},\ }\href
  {\doibase 10.1021/nn3047605} {\bibfield  {journal} {\bibinfo  {journal} {ACS
  Nano}\ }\textbf {\bibinfo {volume} {6}},\ \bibinfo {pages} {11059} (\bibinfo
  {year} {2012})},\ \bibinfo {note} {pMID: 23186074},\ \BibitemShut {NoStop}%
\bibitem [{\citenamefont {Murata}\ \emph {et~al.}(2016)\citenamefont {Murata},
  \citenamefont {Asakawa}, \citenamefont {Nagashima}, \citenamefont
  {Furukawa},\ and\ \citenamefont {Sazaki}}]{murata16}%
  \BibitemOpen
  \bibfield  {author} {\bibinfo {author} {\bibfnamefont {K.-i.}\ \bibnamefont
  {Murata}}, \bibinfo {author} {\bibfnamefont {H.}~\bibnamefont {Asakawa}},
  \bibinfo {author} {\bibfnamefont {K.}~\bibnamefont {Nagashima}}, \bibinfo
  {author} {\bibfnamefont {Y.}~\bibnamefont {Furukawa}}, \ and\ \bibinfo
  {author} {\bibfnamefont {G.}~\bibnamefont {Sazaki}},\ }\href@noop {}
  {\bibfield  {journal} {\bibinfo  {journal} {Proc. Nat. Acad. Sci.}\ }\textbf
  {\bibinfo {volume} {113}},\ \bibinfo {pages} {E6741} (\bibinfo {year}
  {2016})}\BibitemShut {NoStop}%
\end{thebibliography}%

\clearpage

\appendix

\onecolumngrid

\setcounter{page}{1}
\pagenumbering{arabic}

{\centering
{\large Supporting Information for}

{\Large Surface Van der Waals Forces in a Nutshell
\\ by \\}
{\large Luis G. MacDowell}

{\normalsize
Departamento de Qu\'{\i}mica F\'{\i}sica, Facultad de Ciencias
Qu\'{\i}micas, Universidad Complutense, Madrid, 28040, Spain.}

}

\vspace*{1cm}

This document contains supporting information on the derivation of 
results from the main paper. To facilitate cross referencing, this
materials is written as  an appendix section. The 
equation numbering and bibliography  follow the original paper, with
equation labels and references
not in this document  referring to those of the original paper.


\section{DLP theory for the Hamaker constant}

The exact result for the Hamaker function from DLP theory is:
\begin{equation}\label{eq:hamaker}
  A(\h) = - \frac{3}{2} k_B T \sum_{n=0}^{\infty}\, ' 
\int_{r_n}^{\infty} x
\ln([ 1 -  R^M_{1m2}(x,n) e^{-x}] [ 1 - R^E_{1m2}(x,n) e^{-x}]
   )
  dx
\end{equation}
with $R^{M}_{1m2}(x,n)=\Delta^M_{1m}(x,n)\Delta^M_{2m}(x,n)$,
$R^{E}_{1m2}(x,n)=\Delta^E_{1m}(x,n)\Delta^E_{2m}(x,n)$, and
\begin{equation}
\begin{array}{lcl}
 \Delta^M_{ij} = \frac{x_i \epsilon_j - x_j\epsilon_i}{x_i \epsilon_j +
x_j\epsilon_i}
 & &
 \Delta^E_{ij} = \frac{x_i \mu_j - x_j\mu_i}{x_i \mu_j + x_j\mu_i}
\end{array}
\end{equation}
while
\begin{equation}
x_i^2 = x^2 + (\epsilon_i\mu_i - \epsilon_m\mu_m) (2  \omega_n \h/c)^2
\end{equation}
where $\mu_i$ are magnetic permitivities and the rest of the notation
as  described in the main text. 

In practice, all analytical approximations assume
$R^{M}_{1m2}$ and $R^{E}_{1m2}$ are sufficiently smaller
than unity that the logarithms can be Taylor expanded to first order.
Furthermore, without any significant loss in accuracy one can also assume that
the magnetic permitivities amount to unity in all three media.
With these approximations, the above result becomes the first order
approximation of \Eq{1stol}. In our practice, this approximation
is extremely accurate except for $\h\to 0$, where next to leading
terms in the expansion of the logarithm become relevant and can account for
an additioan  5\% contribution to the Hamaker constant.

\section{The Gaussian quadrature method}

An n-point Gaussian quadrature rule is a numerical 
method where one seeks an approximation for
the integral:
\begin{equation}\label{eq:integral}
   I = \int_a^b f(x) g(x) dx
\end{equation} 
such that 
\begin{equation}
   I \approx \sum_i^n f(x_i) m_i
\end{equation} 
The set of points $\{x_i\}$ and weights $\{m_i\}$
are selected such that the integrals
\begin{equation}
   I_i = \int_a^b x^i g(x) dx
\end{equation} 
up to $i=2n-1$ are exact. This guarantees tha the n-point
method is exact for all polynomial $f(x)$ of
degree $2n-1$ or less.

\section{Problem adapted one point Gauss quadrature as an analytical tool}

In his book on van der Waals forces, \citet{parsegian06} describes numerical
methods for the calculation of the Hamaker function 
and strongly advocates the use of the Gaussian quadrature,
claiming that evaluation of the integrand with just a few points
is often sufficient for an accurate calculation. In our paper
we pushed this claim to the maximum, by developing 
problem adapted one point
quadrature rules. We have found this provides an extremely
powerful general mathematical method for analytical calculations.

Accordingly, we attempt to evaluate the integral \Eq{integral} as:
\begin{equation}\label{eq:rule}
    I = f(x^*) m
\end{equation} 
To implement the one-point quadrature rule we calculate:
\begin{equation}\label{eq:integrals}
   I_0 = \int_a^b g(x) dx
\end{equation}  
and
\begin{equation}
   I_1 = \int_a^b x g(x) dx
\end{equation}  
exactly for the weight function $g(x)$.
Then, according to the quadrature rule \Eq{rule},
\begin{equation}
\begin{array}{ccc}
    I_0 & = & m \\
    I_1 & = & x^* m
\end{array}
\end{equation} 
So that trivially it follows
\begin{equation}
\begin{array}{ccc}\label{eq:1pgq}
   m & = & I_0 \\
   x^* & = & \frac{I_1}{I_0} 
\end{array}
\end{equation} 
We find that this is an extremely powerful analytical tool
provided that the weight functions $g(x)$ and interval
of integration are chosen consistent with the physics of the 
problem in hand. i.e., in such a way that 1) the integral
of $g(x)$ converges in the interval $[a,b]$ and 2) that
the remaing function $f(x)$ is only weakly dependent on $x$.

\section{Quadrature for \protect\Eq{1stol}}

In \Eq{1stol} we seek to solve the integral 
\begin{equation}
 I = \int_{r_n}^{\infty} R(x) x e^{-x} dx
\end{equation} 
with  $R(x)$  a bounded and smooth function. The convergence of
the integral is therefore dictated by the term $x e^{-x}$. Accordingly, 
we choose
$f(x)=R(x)$ and the weight function $g(x)=x e^{-x}$.
Applying the one point quadrature rule, \Eq{rule} we find:
\begin{equation}
  I \approx R(x^*) m
\end{equation} 
For the special case  $r_n=0$, this corresponds to
a generalized Gauss-Laguerre one-point quadrature rule. However, using
the correct limit of integration is essential to describe
the correct behavior. Therefore, we develop the rule adapted
to match the physics of the problem. Using \Eq{integrals} and
the range of integration between $r_n$ and $\infty$, the required 
integrals are trivially calculated and yield:
\begin{equation}
   I_0 = ( 1 + r_n ) e^{-r_n}
\end{equation} 
\begin{equation}
   I_1 = ( 2 + 2 r_n + r_n^2 ) e^{-r_n}
\end{equation} 
Using this results, together with \Eq{1pgq}, yields
Eq.8 and Eq.9 of the paper.

\section{Quadrature for \protect\Eq{trivial}}

In \Eq{trivial} we seek to solve the integral 
\begin{equation}
 I = \int_{\nu_T}^{\infty} R(x) [1 + \h x] e^{-\h x} dx
\end{equation} 
Here, the convergence is dictated by $[1 + \h x] e^{-\h x}$ for
large $\h$, but it is dictated by $R(x)$ for $\h\to 0$. Accordingly,
the  natural weighting function for arbitrary $\h$ is
the  full integrand. The corresponding integral cannot be given
in general form for arbitrary dielectric functions. However,
noticing that  $R(x)$  is a smoothly
convergent function that falls off for  $x$ larger than a material
parameter $\nu_{\infty}$, we write:
\begin{equation}
 I = \int_{\nu_T}^{\infty} R(x)   e^{x/\nu_{\infty}}  
\left [ e^{-x/\nu_{\infty}} [1 + \h x] e^{-\h x} \right ] dx
\end{equation} 
With this device, the term in square brackets 
converges for arbitrary $\h$, and is dominated by
the $[1 + \h x] e^{-\h x}$ for large $\h$ as in the original
integrand.

We can now apply the one-point quadrature rule \Eq{rule}
for the choice
$f(x)=R(x) e^{x/\nu_{\infty}}$ and the weight function $g(x)=e^{-x/\nu_{\infty}}[1+\h x] e^{-\h x}$,
so that:
\begin{equation}
  I \approx R(x^*) e^{x^*/\nu_{\infty}} m
\end{equation} 
For the special case  $\nu_T=0$, the resulting quadrature corresponds
to the sum of two generalized Gauss-Laguerre quadrature rules.
Again, considering the lower integration limit is essential to retain
the physics, so we develop a generalized Gauss-Laguerre quadrature with
finite lower bound using \Eq{integrals}. The required integrals are trivially calculated and yield:
\begin{equation}
   I_0 =\frac{1}{\nu_{\infty}} \frac{( \nu_T\h + 1 )(\nu_{\infty}\h + 1) + \nu_{\infty} h }
               { (\nu_{\infty}\h + 1)^2    } 
        e^{-\nu_T\h - \frac{\nu_T}{\nu_{\infty}}}
\end{equation} 
\begin{equation}
   I_1 = \frac{1}{\nu_{\infty}} \frac{( \nu_T\h + 1 )(\nu_{\infty}\h + 1)^2\nu_T
    + (2\nu_T\h + 1 )(\nu_{\infty}\h + 1)\nu_{\infty} + 2\nu_{\infty}^2 h }
               { (\nu_{\infty}\h + 1)^3    } 
        e^{-\nu_T\h - \frac{\nu_T}{\nu_{\infty}}}
\end{equation} 
Using these results together with \Eq{1pgq} yield \Eq{gauss-lifshitz} and
\Eq{nustar} of the paper.

\section{Application of the Euler-McLaurin formula}

The Euler-McLaurin formula allows to relate sums to integrals in exact form, and
thus provides an estimate of the error that results from approximating a
sum by an integral, as in the transformation from \Eq{aproxsum} to \Eq{ahigh1}.

The formula states:
\begin{equation}
\sum_{n=a}^{b} f(n) = \int_a^b f(n) \, dn + \frac{1}{2}(f(a)+f(b)) +
  \sum_{k=1}^{\infty} \frac{B_{2k}}{2k!} ( f^{(2k-1)}(b) - f^{(2k-1)}(a) )
\end{equation} 
where $B_i$ are the Bernoulli coefficients, and $f^{(m)}$ stands for $f$'s m-th
derivative. Applying this formula to order $k=1$ to the sum of \Eq{aproxsum} 
we can obtain next to leading corrections, $\Delta A$, to the integral 
\Eq{ahigh1}, with:
\begin{equation}\label{eq:euler-mclaurin}
 \Delta A = \frac{3}{2}k_BT \left[
   \frac{1}{2} R(x^*,\nu_T) (1+\nu_T\h ) e^{-\nu_T \h} - 
   \frac{1}{12} R(x^*,\nu_T) (\nu_T\h)^2  e^{-\nu_T \h}  
  \right ]
\end{equation} 
where we have assumed that $\epsilon_m$ is constant for the infrared
frequency $\omega_T$, such that $j_m=1$. 

In the limit $\h\to 0$, the correction to the Hamaker constant is
of order $k_BT$, which is a small fraction of the exact
result of order $\hbar\omega_e$.  Only for $\h\nu_T\gg 1$ does the corrections 
become  of the same order as the  integral. However, in that limit,
the result obtained by integration changes only by a multiplicative 
factor of order unity. i.e., the qualitative behavior sought here remains 
unchanged.

The Euler-McLaurin corrections are particularly convenient in
the framework of the FC approximation, since the corrections
can be taken into 
account explicitely without any significant increasee in
algebraic complexity. Indeed, integration of \Eq{2ndmvt} yields:
\begin{equation}\label{eq:simple}
A_{\omega>0}(\h) =  \frac{3 \hbar c}{8\pi\h}
 \tilde{R}^*_{\xi=0}
 [(2+\nu_T \h) e^{-\nu_T \h} - (2 + \nu_{\infty} \h )
e^{-\nu_{\infty} \h}]
\end{equation} 
While adding the correction $\Delta A$ to this expression now provides:
\begin{equation}
A_{\omega>0} =  \frac{3 \hbar c}{8\pi\h}
 \tilde{R}^*_{\xi=0}
 [
 (2+\frac{3}{2}\nu_T \h+\frac{1}{2}(\nu_T\h)^2-\frac{1}{12}(\nu_T\h)^3) 
   e^{-\nu_T \h} - (2 + \nu_{\infty} \h )
e^{-\nu_{\infty} \h}]
\end{equation} 
whence, retaining only the zeroth order correction yields an improved
expression whith  exactly the same algebraic complexity as \Eq{simple}.

\section{Asymptotic expansion of $A_{\omega>0}$(\h) for $\h\to 0$.}

In this section we show that the exact expansion of the Hamaker function
for small $\h$ features next to leading order terms exhibiting
the logarithmic singularity of $O(\h^2\ln\h)$.

To show this we first notice the small $\h$ expansion of $\Delta_{im}$ 
corresponds to  an expansion in powers of $r_n$ about $r_n=0$.
Accordingly, we find:
\begin{equation}
  \Delta_{im} = \frac{\epsilon_i-\epsilon_m}{\epsilon_i+\epsilon_m}
  + \frac{(\epsilon_i-\epsilon_m)\epsilon_i}{(\epsilon_i+\epsilon_m)^2}
   \frac{r_n^2}{x^2}
\end{equation} 
Notice the variable $x$ appears as  $x^{-2}$ in the second order
term of the expansion. Upon integration, this will result in  
the logarithmic singularity mentioned before.  
Using the above result for $\Delta_{im}$, we can now write:
\begin{equation}
  R_{1m2}^M(x,n) = R_0 + R_2 \frac{r_n^2}{x^2}
\end{equation} 
with 
\begin{equation}
     R_0 =  \frac{\epsilon_1-\epsilon_m}{\epsilon_1+\epsilon_m}
            \frac{\epsilon_2-\epsilon_m}{\epsilon_2+\epsilon_m}
\end{equation} 
and
\begin{equation}
   R_2 =  \frac{(\epsilon_1-\epsilon_m)(\epsilon_2-\epsilon_m)\epsilon_2}
               {(\epsilon_1+\epsilon_m)(\epsilon_2-\epsilon_m)^2}
                +
          \frac{(\epsilon_1-\epsilon_m)(\epsilon_2-\epsilon_m)\epsilon_1}
               {(\epsilon_2+\epsilon_m)(\epsilon_1-\epsilon_m)^2}
\end{equation} 
Performing the integration over $x$, we find:
\begin{equation}
 \int_{r_n}^{\infty} R_{1m2}^M(x,n) x e^{-x} d x =
   R_0 [ 1 + r_n ] e^{-r_n} + R_2 r_n^2 Ei(r_n)
\end{equation} 
where $Ei(x)$ is the exponential integral.
We find that right beyond the zero order contribution, the problem
becomes plagued with the non-elementary exponential integral. 
Expanding this result for small $\h$,
we obtain the lowest order correction to the
Hamaker function as:
\begin{equation}
    A_{\omega>0}(\h) = A_{\omega>0}(0) - \frac{3}{2}k_BT
  \sum_{n=1}^{\infty} \left\{\frac{1}{2} R_0 + [\gamma+\ln(\nu_n\h)]R_2 \right \} (\nu_n\h)^2
\end{equation} 
where $\nu$ is defined here as in the paper and $\gamma$ is the Euler constant.
Notice that this small $\h$ expansion does not bare any contribution
from $R_{1m2}^E(x,n)$, which only provides terms of order $\h^4$ or higher.

\section{Second mean value theorem of definite integrals}

According to the second mean value theorem, if $f(x)$ is a monotonic
function and $g(x)$ is an integrable function in the interval
$(a,b)$, there exists a constant $a<c<b$, such that:
\begin{equation}
  \int_a^b f(x) g(x) dx = f(a^+)\int_a^c g(x) dx + f(b^-) \int_c^b g(x) dx
\end{equation} 
This statement is also true when $b\to\infty$, provided that $f(x)$ and
$g(x)$ are bounded functions in the interval $(a,\infty)$.

In order to apply the theorem to solve \Eq{ahigh1}, we consider $x=\nu$, $a=\nu_T$,
$b=\infty$,
 $f(\nu)=\tilde R(x^*,\nu)$ and $g(\nu)=(1+\nu\h) e^{-\nu\h}$. Then, we find:
\begin{equation}
A_{\omega>0}(\h) = \frac{3 \hbar c}{8\pi}
  \tilde{R}(x^*,\nu_T)
\int_{\nu_T}^{\nu_{\infty}}
 (1 + \nu \h)  e^{-\nu \h} d\nu
\end{equation}
This theorem is physically appealing as it implies the existence of a natural
ultraviolet cutoff for the improper integral of \Eq{ahigh1}.  Performing the remaining 
trivial integration yields the result of \Eq{fca}.



\section{Calculation of the high frequency cutoff}

The high frequency cutoff, $\nu_{\infty}$ is calculated by matching
the approximate expression for $A_{\omega>0}$ in the limit of
$\h\to 0$, with the exact result in that limit.

In the WQ approximation, taking \Eq{gauss-lifshitz} in the
limit of $\h\to 0$ one finds:
\begin{equation}\label{eq:hamaker_wq}
  A_{\omega>0}^{WQ}(\h\to 0) = \frac{3}{2}k_BT  
        \frac{\nu_{\infty}\epsilon_m^{1/2}(i\nu_T)}{\nu_{T}\epsilon_m^{1/2}(i\nu_{\infty})} j_m^{-1}(\nu_{\infty}) 
           R(2,\nu_{\infty})\, e 
\end{equation} 
where we have taken into account that $x^*=2$ for $\h=0$, and we
have assumed $\nu^*=\nu_T+\nu_{\infty}\approx\nu_{\infty}$.
By matching this result with the exact Hamaker constant, a trascendental
equation for $\nu_{\infty}$ is obtained. In practice, it is
far more convenient to rewrite the wave numbers in terms of 
angular frequencies (i.e., $\nu_k = 2 \epsilon^{1/2}(i\nu_k) \omega_k/c$).
Taking this into account, the result simplifies to:
\begin{equation}
  A_{\omega>0}^{WQ}(\h\to 0) =   
 \frac{3 \hbar \omega_{\infty}}{4\pi}	j_m^{-1}(\omega_{\infty})
           R(2,\omega_{\infty})\, e 
\end{equation} 
matching this result with the exact Hamaker constant provides
a trascendental equation for $\omega_{\infty}$.

In the FC approximation, one obtains:
\begin{equation}\label{eq:hamaker_fc}
  A^{FC}_{\omega>0}(\h\to 0) = \frac{3}{2}k_BT  
        \frac{\nu_{\infty}}{\nu_{T}}  j_m^{-1}(\nu_{T})  R(2,\nu_{T}) 
\end{equation} 
where again we took into account that $x^*=2$. 
Matching the above result with 
the exact Hamaker constant, a simple linear equation for  $\nu_{\infty}$ 
is obtained.  

More explicit results may be obtained for specific choices of
the dielectric functions. 

\subsection{Specific results for the damped oscillator model}

In the damped oscillator model, the dielectric function is given as:
\begin{equation}
 \epsilon(i\omega) = 1 + \frac{n^2-1}{1+(\omega/\omega_e)^2}
\end{equation} 

\subsubsection{Hamaker constant}

The exact Hamaker constant in the linear approximation of \Eq{1stol}, is
given as:
\begin{equation}
  A_{\omega>0}(\h=0) = \frac{3}{2}k_BT\sum_{n=1}^{\infty} 
  \left (\frac{\epsilon_1-\epsilon_m}{\epsilon_1+\epsilon_m} \right)
  \left (\frac{\epsilon_2-\epsilon_m}{\epsilon_2+\epsilon_m} \right) 
\end{equation} 
where we have taken into account that in this limit ($\h\to 0$),
$R_{1m2}^E$ vanishes exactly, while  $R_{1m2}^M$ becomes independent of $x$, so
that the integration over $x$ may be performed trivially. Conventionally,
in this approximation the sum is now replaced by an integral,
while the three materials are assumed to have the same characteristic
high energy absorbtion frequency $\omega_e$. A 
second integration over the fequencies then provides
the known result:
\begin{equation}\label{eq:tw}
  A_{\omega>0}(\h=0) = \frac{3\hbar \omega_e}{8\sqrt{2}}
  \frac{(n_1^2-n_m^2)(n_2^2-n_m^2)}{(n_1^2+n_m^2)^{1/2}(n_2^2+n_m^2)^{1/2}
          [(n_1^2+n_m^2)^{1/2}+(n_2^2+n_m^2)^{1/2}]}
\end{equation} 
We use below this result as the 'exact' Hamaker constant.

\subsubsection{WQ approximation}

In the WQ approximation, the damped oscillator model for the dielectric 
functions yields:
\begin{equation}
           R(2,\omega_{\infty}) = 
\left(\frac{\omega_e^2(n_1^2-n_m^2)}{\omega_e^2(n_1^2+ n_m^2)+2\omega_{\infty}^2}\right)
\left(\frac{\omega_e^2(n_2^2-n_m^2)}{\omega_e^2(n_2^2+ n_m^2)+2\omega_{\infty}^2}\right)
\end{equation} 
Replacing this result into \Eq{hamaker_wq}, matching the resultant approximate
Hamaker constant with that of \Eq{tw} and introducing the factor $f$
such that $\omega_{\infty}=f(n_1+n_m^2)^{1/4}(n_2^2+n_m^2)^{1/4}\omega_e$, 
we obtain:
\begin{equation}
 \left(1 + \frac{1}{2} \frac{d\ln\epsilon_m}{d\ln\omega}\right )^{-1}
  \frac{e\, q^2 f}{(1+2 q^2 f^2)(q^2+2f^2)} = \frac{\pi}{2\sqrt{2}}
  \frac{q}{q^2+1}
\end{equation} 
where $q=(n_1^2+n_m^2)^{1/4}/(n_2^2+n_m^2)^{1/4}$.
This is a complicated  trascendental equation for $f$,
but simplifies very much when the
interactions between material 1 and 2 take place across vacuum,
such that $\epsilon_m$ is a constant. For the additional simplification
that materials 1 and 2 are the same, $q=1$, one obtains
$f=0.5934$ 

\subsubsection{FC approximation}

In the FC approximation, the damped oscillator model for the dielectric 
functions yields:
\begin{equation}
           R(2,\omega_{T}) = 
  \left (\frac{n_1^2-n_m^2}{n_1^2+ n_m^2}\right )
  \left (\frac{n_2^2-n_m^2}{n_2^2+ n_m^2}\right )
\end{equation} 
where we have assumed that $\omega_T/\omega_e\ll 1$.
Replacing this result into \Eq{hamaker_fc} and matching the resulting 
approximate
Hamaker constant with that of \Eq{tw}, yields:
\begin{equation}
\nu_{\infty}=f(n_1^2+n_m^2)^{1/4}(n_2^2+n_m^2)^{1/4}n_m \frac{\omega_e}{c} 
\end{equation} 
with
\begin{equation}
   f = \frac{\pi}{\sqrt{2}} \frac{q}{q^2+1}
\end{equation} 
where we have taken into accout that for a single
oscillator in the UV region, $j_m=1$ at the thermal frequency $\omega_T$.
Alternatively, the above result can be given
explictely in terms of the refraction indexes as:
\begin{equation}
  \nu_{\infty}= \frac{\pi}{\sqrt{2}} n_m 
  \frac{(n_1^2+n_m^2)^{1/2}(n_2^2+n_m^2)^{1/2}}
          {(n_1^2+n_m^2)^{1/2}+(n_2^2+n_m^2)^{1/2}} \frac{\omega_e}{c}
\end{equation}

\section{Working formulae in the FCA-t approximation}

For ease of implementation we provide here the final working formula
in the FCA-t approximation:
\begin{equation}
 A_{\omega > 0}(\h) = \frac{3\pi k_BT}{8\sqrt{2}\nu_T\h}      
       \left ( \frac{n_1^2-n_m^2}{n_1^2+n_m^2} \frac{n_2^2-n_m^2}{n_2^2+n_m^2} \right )
       \left [
 (2+\frac{3}{2}\nu_T \h) e^{-\nu_T \h} - (2 + \nu_{\infty} \h ) e^{-\nu_{\infty} \h}
       \right ]
\end{equation} 
or
\begin{equation}
 A_{\omega > 0}(\h) = \frac{3\hbar c}{32\sqrt{2}\,n_m\h}      
       \left ( \frac{n_1^2-n_m^2}{n_1^2+n_m^2} \frac{n_2^2-n_m^2}{n_2^2+n_m^2} \right )
       \left [
 (2+\frac{3}{2}\nu_T \h) e^{-\nu_T \h} - (2 + \nu_{\infty} \h ) e^{-\nu_{\infty} \h}
       \right ]
\end{equation} 
where the wave-numbers $\nu_T$ and $\nu_{\infty}$ are:
\begin{equation}
\left \{
\begin{array}{ccc}
   \nu_T & = & \frac{4\pi k_B T n_m}{c \hbar } \\
         &   &  \\
   \nu_{\infty} & = &  4 n_m  
  \frac{(n_1^2+n_m^2)^{1/2}(n_2^2+n_m^2)^{1/2}}
          {(n_1^2+n_m^2)^{1/2}+(n_2^2+n_m^2)^{1/2}} \frac{\omega_e}{c}
\end{array}
\right .
\end{equation} 

\section{Material parameters for the oscillator model}

\subsection{mica/air/mica}

For the  interaction of two mica plates across air we used data for muscovite
mica described using the dielectric response 
given by \Eq{oscillator}, with parameters obtained from high
frequency data by \citet{bergstrom97}, i.e.,
$n^2=2.508$ and $\omega_e=1.963\cdot 10^{-16}$~rad/s. 
The static permitivity is $\epsilon(0)=5.4$.

\subsection{mica/water/mica}

For the interaction of mica across air we employ the one oscillator model
with data as summarized in \citet{israelachvili91}. Parameters for water are $\epsilon(0)=80$,
$n^2=1.7769$ and $\omega_e=1.885\cdot 10^{-16}$~rad/s. Parameters for mica are
$\epsilon(0)=7.0$,  $n^2=2.56$ and $\omega_e=1.885\cdot 10^{-16}$~rad/s.

\subsection{octane/water/air}

For the interaction of a water film adsorbed on the octane/air interface we
use parameters for water as described above, and the parameters for octane
as given in \citet{israelachvili91}, i.e., 
$\epsilon(0)=1.95$, $n^2=1.9238$ and $\omega_e=1.885\cdot 10^{-16}$.

\end{document}